\documentclass[twocolumn,showpacs,amsmath,amssymb,10pt]{revtex4}
\usepackage{graphicx,color}
\usepackage{amsmath}
\usepackage{amssymb}
\usepackage{bbm}

\usepackage{amsfonts}
\usepackage{bm}

\newcommand{\ud}{\mathrm{d}}
\newcommand{\ot}{{\,\otimes\,}}
\newcommand{{\Cd}}{{\mathbb{C}^d}}

\newcommand{\tr}{\mathrm{Tr}}
\newcommand{\bra}[1]{\mbox{$\langle #1 |$}}

\newcommand{\ket}[1]{\mbox{$| #1 \rangle$}}
\def\oper{{\mathchoice{\rm 1\mskip-4mu l}{\rm 1\mskip-4mu l}
{\rm 1\mskip-4.5mu l}{\rm 1\mskip-5mu l}}}
\def\<{\langle}
\def\>{\rangle}

\newtheorem{Example}{Example}
\newtheorem{Proposition}{Proposition}

\newcommand{\beq}{\begin{equation}}
\newcommand{\eeq}{\end{equation}}
\newcommand{\bear}{\begin{eqnarray}}
\newcommand{\ear}{\end{eqnarray}}
\newcommand{\bdm}{\begin{displaymath}}
\newcommand{\edm}{\end{displaymath}}
\newcommand{\sz}{\sigma_z}
\newcommand{\sx}{\sigma_x}
\newcommand{\sy}{\sigma_y}

\begin{document}
\title{\textbf{
Non-Markovian random unitary qubit dynamics}}
\author{Dariusz Chru\'sci\'nski and Filip A. Wudarski }
\affiliation{ Institute of Physics \\ Faculty of Physics, Astronomy and Informatics, Nicolaus Copernicus University \\
Grudzi{a}dzka 5, 87--100 Torun, Poland\\
}

\begin{abstract}
We compare two approaches to non-Markovian quantum evolution: one based on the concept of divisible maps and the other one based on distinguishability of quantum states. The former concept is fully characterized in terms of local generator whereas it is in general not true for the latter one. A simple example of random unitary dynamics of a qubit shows the intricate difference between those approaches. Moreover, in this case both approaches are fully characterized in terms of local decoherence rates. As a byproduct it is shown that entropy might monotonically increase even for non-Markovian qubit dynamics.


\end{abstract}


\maketitle

Quantum dynamics is represented by the dynamical map, that is, a family of completely positive and trace preserving maps $\Lambda_t$ $(t\geq 0)$ such that $\Lambda_0 = \oper$. If $\rho$ is an initial state of the system then $\rho_t = \Lambda_t(\rho)$ defines evolution of $\rho$. This description provides generalization of unitary evolution $\Lambda_t(\rho) = U_t \rho U_t^\dagger$ with unitary $U_t = e^{-iHt}$, where $H$ represents the Hamiltonian of the system. Any departure from unitary evolution signals the nontrivial interaction of the system with an environment which is responsible for decoherence and dissipation processes \cite{Breuer} in open quantum system. The traditional approach to the dynamics of such systems consists in applying a suitable Born-Markov approximation leading to the celebrated
quantum Markov semigroup \cite{GKS,Lindblad,Alicki} which neglects all memory effects. Recent theoretical activity and technological progress  show the importance of more refine approach based on non-Markovian evolution. Non-Markovian quantum dynamics becomes in recent years very active field of both theoretical and experimental research (see e.g. recent papers \cite{Strunz1}--\cite{Franco} and references therein).

We shall call quantum evolution Markovian if the corresponding dynamical map $\Lambda_t$ is divisible \cite{RHP}. Let us recall
that $\Lambda_t$ is divisible if $\Lambda_t = V_{t,s} \Lambda_s$ and $V_{t,s}$ is completely positive and trace preserving for all $t\geq s$, that is, it gives rise to 2-parameter family of legitimate propagators. The essential property of $V_{t,s}$ is the following (inhomogeneous) composition law
 $   V_{t,s} \, V_{s,u} = V_{t,u}$
for all $t\geq s\geq u$. It should be stressed that Markovian dynamics (divisible map) is entirely characterized by the properties of the local in time generators $L_t$, that is, if $\Lambda_t$ satisfies $\dot{\Lambda}_t =L_t \Lambda_t$, then $\Lambda_t$ corresponds to Markovian dynamics if and only if $L_t$ has the standard form \cite{GKS,Lindblad} for all $t\geq 0$, that is,
\begin{equation*}\label{}
    L_t \rho = -i[H(t),\rho]  + \sum_\alpha \left( V_\alpha(t) \rho V_\alpha^\dagger(t) - \frac 12 \{ V_\alpha^\dagger(t)V_\alpha(t),\rho\} \right)  ,
\end{equation*}
with time dependent Hamiltonian $H(t)$ and noise operators $V_\alpha(t)$. A different approach to Markovianity was proposed by Breuer et. al. \cite{BLP}: authors of \cite{BLP} define non-Markovian dynamics as
a time evolution for the open system characterized by a temporary
flow of information from the environment back into the system and
manifests itself as an increase in the distinguishability of pairs
of evolving quantum states:
\begin{equation}\label{BLP-c}
    \sigma(\rho_1,\rho_2;t) = \frac 12 \frac{d}{dt} \, ||\Lambda_t(\rho_1-\rho_2)||_1\ ,
\end{equation}
where $||A||_1 = {\rm Tr}\sqrt{A^\dagger A}$ denotes the trace norm. According to \cite{BLP} the dynamics $\Lambda_t$ is markovian iff $ \sigma(\rho_1,\rho_2;t) \leq 0$ for all pairs of states $\rho_1,\rho_2$ and $t \geq 0$. Contrary to divisible map this approach does not correspond to any composition law and is not characterized by the properties of local generator. It turns out that condition $\sigma(\rho_1,\rho_2;t) \leq 0$ is less restrictive than requirement of complete positivity for $V_{t,s}$ and one can construct  $\Lambda_t$ which is non-Markovian (not divisible) but still  gives rise to the negative flow of information (see \cite{versus1,versus2,versus3}).

In this report we provide a simple example of qubit dynamics for which the condition $ \sigma(\rho_1,\rho_2;t) \leq 0$ is fully characterized by local generator. Consider the following random unitary dynamical map
\begin{equation}\label{Pauli}
    \Lambda_t\rho = \sum_{\alpha=0}^3 p_\alpha(t) \sigma_\alpha \rho\, \sigma_\alpha\ ,
\end{equation}
where $\sigma_0= \mathbb{I}$, and $\sigma_1,\sigma_2,\sigma_3$ are Pauli matrices, and $p_\alpha(t)$ is time-dependent probability distribution such that $p_0(0)=1$ which guarantee that $\Lambda_0=\oper$. This map was recently investigated in \cite{Bassano}. It is clear that one may consider (\ref{Pauli}) as a family of Pauli channels.
To answer the question wether $\Lambda_t$ is Markovian one has to analyze the corresponding local generator. To find  $L_t = \dot{\Lambda}_t \Lambda_t^{-1}$ one has to compute the inverse map $\Lambda_t^{-1}$. Let us observe that
\begin{equation}\label{}
    \Lambda_t(\sigma_\alpha) = \lambda_\alpha(t) \sigma_\alpha\ ,
\end{equation}
where the time-dependent eigenvalues are given by
\begin{equation}\label{}
    \lambda_\alpha(t) = \sum_{\beta=0}^3 H_{\alpha\beta} \, p_\beta(t)\ ,
\end{equation}
with $H_{\alpha\beta}$ being the Hadamard matrix
\begin{equation}\label{}
    H = \left( \begin{array}{rrrr} 1 & 1 & 1 & 1 \\
    1 & 1 & -1 & -1 \\
    1 & -1 & 1 & - 1 \\
    1 & -1 & 1 & 1 \end{array} \right) \ .
\end{equation}
Note that $\lambda_0(t)=1$ and $|\lambda_k(t)|\leq 1$ for $k=1,2,3$. It is therefore clear that
\begin{equation}\label{}
    L_t(\sigma_\alpha) = \mu_\alpha(t) \sigma_\alpha\ ,
\end{equation}
where
\begin{equation}\label{}
    \mu_\alpha(t) = \frac{\dot{\lambda}_\alpha(t)}{\lambda_\alpha(t)}\ ,
\end{equation}
and hence in particular $\mu_0(t)=0$ due to $\lambda_0(t)=1$. Introducing
\begin{equation}\label{}
    \gamma_\alpha(t) = \frac 14 \sum_{\beta=0}^3 H_{\alpha\beta} \, \mu_\beta(t)\ ,
\end{equation}
the local generator $L_t$ is given by
\begin{equation}\label{}
    L_t(\rho) =  \sum_{\alpha=0}^3 \gamma_\alpha(t) \sigma_\alpha \rho \sigma_\alpha \ .
\end{equation}
Finally, observing that $$\sum_{\alpha=0}^3 \gamma_\alpha(t) = \frac 14 \sum_{\alpha,\beta=0}^3 H_{\alpha\beta} \, \mu_\beta(t) = \mu_0(t)=0 \ ,$$  one arrives at the standard form
\begin{equation}\label{}
    L_t(\rho) = \sum_{k=1}^3 \gamma_k(t) ( \sigma_k \rho \sigma_k - \rho) \ .
\end{equation}
Hence
\begin{Proposition}
The random unitary dynamics (\ref{Pauli}) is Markovian iff
\begin{equation}\label{1gamma}
    \gamma_1(t) \geq 0\ , \ \  \gamma_2(t) \geq 0\ , \ \  \gamma_3(t) \geq 0\ ,
\end{equation}
for all $t \geq 0$.
\end{Proposition}
Note, that  (\ref{1gamma}) yields highly nontrivial conditions for probability distribution $p_\alpha(t)$:
\begin{equation}\label{}
     \sum_{\beta=0}^3 H_{k\beta} \left\{ \frac{  \sum_{\nu=0}^3 H_{\beta\nu} \dot{p}_\nu(t)}{ \sum_{\sigma=0}^3 H_{\beta\sigma} {p}_\sigma(t)} \right\} \, \geq \, 0 \ ,
\end{equation}
for $k=1,2,3$ and $t\geq 0$. The inverse relations yield
\begin{eqnarray*}
p_0(t) &=&\frac{1}{4}\, [1+ A_{12}(t) + A_{13}(t) + A_{23}(t)] \ , \\
p_1(t) &=&\frac{1}{4}\, [1- A_{12}(t) - A_{13}(t) + A_{23}(t)] \ ,\\
p_2(t) &=&\frac{1}{4}\, [1- A_{12}(t) + A_{13}(t) - A_{23}(t)] \ ,\\
p_3(t) &=&\frac{1}{4}\, [1+ A_{12}(t) - A_{13}(t) - A_{23}(t)] \ ,
\end{eqnarray*}
where $A_{ij}(t) = e^{-2[\Gamma_i(t) + \Gamma_j(t)]}$ and we introduced $$\Gamma_k(t) = \int_0^t \gamma_k(\tau) d\tau\ . $$
Let us recall \cite{WIT} that if $\Lambda_t$ is Markovian then
\begin{equation}\label{}
    \frac{d}{dt}\, || \Lambda_t(X) ||_1 \leq 0\ ,
\end{equation}
for all Hermitian $X$. Taking $X = \sigma_k$ one obtains
\begin{equation}\label{}
    \frac{d}{dt}\, || \Lambda_t(\sigma_k) ||_1 = \frac{d}{dt}\,|{\lambda}_k(t)| \, ||\sigma_k||_1 \leq 0\ ,
\end{equation}
and hence Markovianity of random unitary evolution (\ref{Pauli}) implies
\begin{equation}\label{dot-lambda}
   \frac{d}{dt}\, |{\lambda}_k(t)| \leq 0 \ ,
\end{equation}
for $k=1,2,3$. One easily finds the following relations
\begin{eqnarray*}
    \lambda_1(t) &=& e^{-2[\Gamma_2(t) + \Gamma_3(t)]}\ , \\
    \lambda_2(t) &=& e^{-2[\Gamma_1(t) + \Gamma_3(t)]}\ , \\
    \lambda_3(t) &=& e^{-2[\Gamma_1(t) + \Gamma_1(t)]}\ ,
\end{eqnarray*}
and hence (\ref{dot-lambda}) implies
\begin{eqnarray}\label{2gamma}
    \gamma_1(t) + \gamma_2(t) &\geq&  0 \ , \nonumber \\
    \gamma_1(t) + \gamma_3(t) &\geq& 0 \ , \\
    \gamma_2(t) + \gamma_3(t) &\geq& 0 \ , \nonumber
\end{eqnarray}
for all $t\geq 0$. We stress that the above conditions are only necessary but not sufficient for Markovianity: (\ref{1gamma}) imply (\ref{2gamma}) but the converse is of course not true.

Now comes our main result
\begin{Proposition}
The random  unitary dynamics (\ref{Pauli}) satisfies $\sigma(\rho_1,\rho_2;t) \leq 0$ if and only if conditions (\ref{2gamma}) are satisfied.
\end{Proposition}

\noindent Proof: let us consider   $||\Delta_t||_1$, where $\Delta_t=\Lambda_t(\Delta)$ and $\Delta=\rho_1-\rho_2$. Since $\Delta$ is traceless one has $\Delta = \sum_{k=1}^3 x_k \sigma_k$ with real $x_k$. Note that $\Delta^2 = (\sum_{k=1}^3 x^2_k)\, \mathbb{I}$. Now, $\Lambda_t(\Delta) = \sum_{k=1}^3 \lambda_k(t) x_k \sigma_k$ is traceless as well, and hence
\begin{equation}\label{}
    ||\Lambda_t(\Delta)||_1 = {\rm Tr}\, \sqrt{ [\Lambda_t(\Delta)]^2 } =  {\rm Tr}\, [\xi(t) \mathbb{I}] = 2\, \xi(t)\ ,
\end{equation}
where $\xi(t) = \sqrt{ \sum_{k=1}^3 |\lambda_k(t)|^2 x_k^2} \geq 0$. Observe that
\begin{eqnarray*}
  \frac 12 \frac{d}{dt}||\Delta_t||_1 = \frac{d\xi(t)}{dt}\,  =  \frac{1}{2\xi(t)} \sum_{k=1}^3 x^2_k \frac{d}{dt}|\lambda_k(t)|^2 \ .
\end{eqnarray*}
It is therefore clear that $\frac{d}{dt}||\Delta_t||_1 \leq 0$ for all pairs $\rho_1$ and $\rho_2$ if and only if $\frac{d}{dt} |{\lambda}_k(t)| \leq 0$ for $k=1,2,3$ which is equivalent to (\ref{2gamma}).

Consider now a general (not necessarily traceless) Hermitian $X = x_0 \mathbb{I} + \Delta$ with ${\rm Tr}\, \Delta=0$. Due to $\Lambda_t(\mathbb{I}) = \mathbb{I}$ one has $X_t = \Lambda_t(X) = x_0 \mathbb{I} + \Delta_t$. Let $\delta_+(t)$ and $\delta_-(t)$ denote the eigenvalues of $\Delta_t$ such that $\delta_+(t) = - \delta_-(t) \geq 0$. The corresponding eigenvalues of $X_t$ read $x_\pm(t) = x_0 \pm \delta_+(t)$ and hence
\begin{equation*}
    ||X_t||_1 = |x_0 + \delta_+(t)| + |x_0 - \delta_+(t)| .
\end{equation*}
Assuming $x_0 >0 $ (for $x_0<0$ the argument is analogous) one finds
\begin{equation*}
    \frac{d}{dt}\, ||X_t||_1 = \left\{ 1 - \frac{x_0 - \delta_+(t)}{|x_0 - \delta_+(t)|} \right\} \, \frac{d}{dt}\, \delta_+(t) \ ,
\end{equation*}
and hence $\frac{d}{dt}\, ||X_t||_1=0$ if $\delta_+(t) <x_0$, and  $\frac{d}{dt}\, ||X_t||_1= 2\frac{d}{dt}\, \delta_+(t) $ if $\delta_+(t) \geq x_0$. For $x_0=0$ one has $ \frac{d}{dt}\,||\Delta_t||_1 = \frac{d}{dt}\,||X_t||_1 = 2\frac{d}{dt}\, \delta_+(t)\leq 0 $ which finally implies  \begin{equation*}
    \frac{d}{dt}\,||\Lambda_t(X)||_1 \leq 0\ ,
\end{equation*}
for arbitrary Hermitian $X$.

Interestingly, similar result holds for the entropy $S(\rho_t)$. Note that $\Lambda_t(\mathbb{I})=\mathbb{I}$ and hence $S(\rho_t) \geq S(\rho)$. Now, if $\Lambda_t$ is Markovian then $\frac{d}{dt} S(\rho_t) \geq 0$ for all initial states $\rho$.

\begin{Proposition}
If conditions (\ref{2gamma}) are satisfied then $\frac{d}{dt} S(\rho_t) \geq 0$.
\end{Proposition}
To prove it observe that decomposing
\begin{equation}\label{}
    \rho = \frac 12\, \mathbb{I} +  \left( \begin{array}{cc} u & w \\ w^* & -u  \end{array} \right) \ ,
\end{equation}
with $w = a + i b$, one finds for the time evolution
\begin{equation}\label{}
    \rho_t = \frac 12\, \mathbb{I} + e^{-[\Gamma_1(t) + \Gamma_2(t)]} \left( \begin{array}{cc} u & w(t) \\ w^*(t) & -u  \end{array} \right) \ ,
\end{equation}
where  $w(t)  = e^{-\Gamma_3(t)} [ e^{\Gamma_1(t)}\, a + i e^{\Gamma_2(t)} \, b]$.
Let $x_+(t) \geq x_-(t)$ be the corresponding eigenvalues of $\rho_t$
\begin{equation*}
    x_\pm(t) = \frac 12 \pm e^{-[\Gamma_1(t) + \Gamma_2(t)]}\, \sqrt{ u^2 + |w(t)|^2} \ .
\end{equation*}
One has
\begin{equation}\label{}
    \frac{d}{dt} S(\rho_t) = - \dot{x}_+(t) \log \frac{x_+(t)}{x_-(t)}\ ,
\end{equation}
and hence $\frac{d}{dt} S(\rho_t) \geq 0$ if $\dot{x}_+(t) \leq 0$. One finds
\begin{eqnarray*}
  \frac{d}{dt}\, x_+(t) =  \frac{e^{-[\Gamma_1(t) + \Gamma_2(t)]}}{\sqrt{ u^2 + |w(t)|^2}}  \  A(t)\ ,
\end{eqnarray*}
where
\begin{eqnarray*}
    A(t)& =& -[\gamma_1(t) + \gamma_2(t)] u^2  \\ &-&  [\gamma_1(t) + \gamma_3(t)] a^2\, e^{2[\Gamma_2(t) - \Gamma_3(t)]} \\
      &-&   [\gamma_2(t) + \gamma_3(t)]\, b^2\, e^{2[\Gamma_1(t) - \Gamma_3(t)]}\ ,
\end{eqnarray*}
and hence $A(t) \leq 0$ whenever (\ref{2gamma}) is satisfied. It is clear that similar result holds if we replace von Neumann entropy by Renyi
or Tsallis generalized entropies \cite{WIT}.

Consider for  example
\begin{equation}\label{}
    \gamma_1(t) = \gamma_2(t) = 1\ , \ \ \ \gamma_3(t) = \sin t\ .
\end{equation}
This choice is perfectly legitimate since  $\Gamma_1(t)=\Gamma_2(t)=t\geq 0$ and $\Gamma_3(t) = 1-\cos t\geq 0$. The corresponding evolution is non-Markovian due to the fact that $\gamma_3(t)$ can be negative but the conditions (\ref{2gamma}) are satisfied. Hence, even if one time-dependent decoherence rate is periodic and takes strictly negative values both information flow and von Neumann entropy behaves perfectly monotonically as in the case of Markovian dynamics.

Interestingly, if there are at most two decoherence channels, that is, at most two $\gamma_k(t)$ are non-vanishing, then (\ref{2gamma}) imply that $\gamma_k(t) \geq 0$ and hence (\ref{1gamma}) and (\ref{2gamma}) are equivalent. However, the relation between $\gamma_k(t)$ and probability distribution $p_\alpha(t)$ is still quite complicated. If there is only one decoherence channel, say $k$th, that is $\gamma_l(t)=0$ for $l \neq k$, then
\begin{equation*}
    p_0(t) = \frac 12 [1 + e^{-2\Gamma_k(t)}]\, \ \ \   p_k(t) = \frac 12 [1 - e^{-2\Gamma_k(t)}]\ ,
\end{equation*}
and $p_l(t)=0$ for $l \neq k$. In this case
\begin{equation}\label{}
    \gamma_k(t) = \frac{\dot{p}_k(t)}{1-2p_k(t)} = - \frac{\dot{p}_0(t)}{2p_0(t)-1}\ ,
\end{equation}
and hence Markovianity is equivalent to $\dot{p}_0(t) = - \dot{p}_k(t) \leq 0$. This way we can fully characterize (non)Markovianity of the bit-flip ($k=1$), phase-flip  ($k=2$), and phase-damping ($k=3)$ dynamics. If $p_1(t)=p_2(t)=p_3(t) = \frac 13 [1-p_0(t)]$, then
\begin{equation}\label{}
    \Lambda_t = p(t)\oper_2 +[1-p(t)] \Lambda_{\rm D}\ ,
\end{equation}
where $p(t) = \frac 13 [4p_0(t) -1]$ and  $\Lambda_{\rm D}$ is a completely depolarizing channel corresponding to $p_\alpha(t) = \frac 14$. Note that in this case $W_t = (\oper \ot \Lambda_t) P^+_2$ defines a family of 2-qubit Werner states satisfying initial condition $W_0 = P^+_2$ ($P^+_2 = \frac 12 \sum_{i,j} |ii\>\<jj|$ denotes canonical maximally entangled state). Such special qubit channels were recently analyzed in Ref. \cite{HINDU}. One has $\lambda_k(t)=p(t)$ for $k=1,2,3$ and hence $\gamma_k(t) = - \dot{p}(t)/[4p(t)]$. It leads to the following local generator $ L_t(\rho) = - \frac{\dot{p}(t)}{4p(t)}\, \sum_{k=1}^3 ( \sigma_k \rho\, \sigma_k - \rho)$ which may be rewritten as follows
\begin{equation}\label{Dep-2}
    L_t = - \frac{\dot{p}(t)}{p(t)}\, [ \Lambda_D - \oper_2] \ ,
\end{equation}
which shows that the corresponding dynamics is Markovian iff $\frac{d}{dt} p(t) \leq 0$.

It is clear that (\ref{Pauli}) may be generalized for arbitrary $N>2$ as follows
\beq\label{RUM}
\Lambda_t(\rho)=\sum_{k,l=0}^{N-1}p_{kl}(t)U_{kl}\rho U_{kl}^\dag,
\eeq
where $p_{kl}(t)$ is time-dependent probability distribution, and  $U_{kl}$ are generalized spin operators defined by
\beq
U_{kl}=\sum_{m=0}^{N-1} \omega^{mk}\ket{m}\bra{m + l}\ ,
\eeq
with $\omega = e^{2\pi i/N}$ (we add modulo $N$). It is legitimate dynamical map provided $p_{00}(0)=1$ which guarantee that $\Lambda_0 = \oper$.  Using
\beq
U_{kl}U_{rs}=\omega^{ks}U_{k+r,l+s}\ , \ \ \ U_{kl}^\dagger = \omega^{kl} U_{-k,-l}\ ,
\eeq
one finds  $\Lambda_t(U_{ij}) = \lambda_{ij}(t) U_{ij}\,$, with time-dependent eigenvalues
\begin{equation}\label{mu-p}
    \lambda_{ij}(t) = \sum_{k,l=0}^{N-1} p_{kl}(t) \omega^{kj-il}\ .
\end{equation}
Note, that $\lambda_{00}(t)=1$ and $|\lambda_{ij}(t)|\leq 1$ for all $t\geq 0$. Clearly one has $\lambda_{ij}(0) =1$ for all $i,j=0,\ldots,N-1$. Introducing two $N \times N$ matrices $\widehat{\lambda}(t)$ and $\widehat{p}(t)$ with matrix elements $\lambda_{ij}(t)$ and $p_{ij}(t)$, respectively, one can rewrite (\ref{mu-p}) in the following compact form
\begin{equation}\label{}
    \widehat{\lambda}(t) = F\, \widehat{p}(t)^{\rm T} F^\dagger\ ,
\end{equation}
where $F$ denote the matrix of the discrete Fourier transform, that is, $F_{il} = \omega^{-il}$. Using $F^{-1} = \frac 1N F^\dagger$ one finds for the inverse relation
\begin{equation}\label{}
    \widehat{p}(t)^{\rm T} = \frac{1}{N^2}\, F^\dagger \, \widehat{\lambda}(t)\, F\ .
\end{equation}
The corresponding time-dependent local generator $L_t$ reads
\begin{equation}\label{}
    L_t(\rho)={\sum_{i,j}}'\gamma_{ij}(t)\, \big( U_{ij}\rho U_{ij}^\dag-\rho\big)\label{gen3}\ ,
\end{equation}
where $\sum_{i,j}' = \sum_{(i,j) \neq(0,0)}$, and
\begin{equation}\label{}
    \widehat{\gamma}(t)^{\rm T} = \frac{1}{N^2}\, F^\dagger \, \widehat{\mu}(t)\, F\ ,
\end{equation}
and $\mu_{ij}(t)$  are eigenvalues of $L_t$ which are related to $\lambda_{ij}(t)$ via $$\lambda_{ij}(t) = \exp\big[ \int_0^t \mu_{ij}(\tau)d\tau \big] \ . $$
We can summarize

\begin{Proposition}
The random unitary dynamics (\ref{RUM}) is Markovian ($\Lambda_t$ is divisible) if and only if $\gamma_{ij}(t) \geq 0$ for each pair $(i,j)\neq (0,0)$ and $t \geq 0$.
\end{Proposition}

In particular taking $p_{ij}(t) = [1- p_{00}(t)]/[N^2-1]$ for $(i,j) \neq (0,0)$ one finds  $\,\Lambda_t = p(t)\oper_N +[1-p(t)] \Lambda_{\rm D}$, where $p(t) = [N^2 p_{00}(t) -1]/[N^2-1]$ and  $\Lambda_{\rm D}$ is a completely depolarizing channel corresponding to $p_\alpha(t) = {1}/{N^2}$. In this case $W_t = (\oper_N \ot \Lambda_t) P^+_N$ defines a family of isotropic states in $\mathbb{C}^N \ot \mathbb{C}^N$ with initial condition $W_0 = P^+_N$. One has $\lambda_{kl}(t)=p(t)$ for $(k,l)\neq (0,0)$ and hence $\gamma_{kl}(t) = - \dot{p}(t)/[N^2p(t)]$. It leads to the following local generator $ L_t(\rho) = - \frac{\dot{p}(t)}{N^2 p(t)}\, {\sum'_{k,l}} [ U_{kl} \rho\, U_{kl}^\dagger - \rho ]$
which may be rewritten as $\,  L_t = - \frac{\dot{p}(t)}{p(t)}\, [ \Lambda_D - \oper_N]$, which shows that the corresponding dynamics is Markovian iff $\frac{d}{dt} p(t) \leq 0$. For $N=2$ it reduces to (\ref{Dep-2}).

It is clear \cite{WIT} that if $\Lambda_t$ is Markovian then
\begin{equation}\label{}
    \frac{d}{dt} |\lambda_{ij}(t)|\leq 0\ ,
\end{equation}
for $i,j=1,\ldots,N-1$. Clearly, one obtains a nontrivial condition only for $(i,j)\neq (0,0)$. Using $\sum_{m,n=0}^{N-1}  \gamma_{mn}(t) =0$, the above formula imply the following conditions for the time-dependent decoherence rates
\begin{equation}\label{Ngamma}
    {\sum_{m,n}}' a_{mn;ij}\ \gamma_{mn}(t) \geq 0\ ,
\end{equation}
where $a_{mn;ij} = 1- {\rm Re}\, \omega^{jn-im}$. For $N=2$ these conditions reduce to (\ref{2gamma}). If $n=3$ the formula (\ref{Ngamma}) provides  the following set of conditions
\begin{eqnarray*}
\gamma_{10}(t) + \gamma_{11}(t) + \gamma_{12}(t) + \gamma_{20}(t)  + \gamma_{21}(t) + \gamma_{22}(t) & \geq & 0 \ , \\
\gamma_{01}(t) + \gamma_{02}(t) + \gamma_{11}(t) + \gamma_{12}(t)  + \gamma_{21}(t) + \gamma_{22}(t) & \geq & 0 \ , \\
\gamma_{01}(t) + \gamma_{02}(t) + \gamma_{10}(t) + \gamma_{12}(t)  + \gamma_{20}(t) + \gamma_{21}(t) & \geq & 0 \ , \\
\gamma_{01}(t) + \gamma_{02}(t) + \gamma_{10}(t) + \gamma_{11}(t)  + \gamma_{20}(t) + \gamma_{22}(t) & \geq & 0 \ .
\end{eqnarray*}
These are of course much more difficult to handle than simple condition (\ref{2gamma}). However, preliminary numerical analysis shows that these conditions might be also sufficient for the negative information flow.

In conclusion we provide a simple example of qubit random unitary dynamics (\ref{Pauli}) and show that two concepts of Markovian dynamics based on divisible dynamical maps and negative flow of information diverge. Interestingly both concepts are fully characterized in terms of local generator: divisibility by (\ref{1gamma}) and negative flow of information by (\ref{2gamma}). We stress that it is the first example of quantum dynamics which provide full characterization  of the Breuer et. al. condition \cite{BLP} in terms of local in time generator $L_t$. Interestingly, condition (\ref{2gamma}) is also sufficient for the monotonicity of von Neumann entropy $\frac{d}{dt} S(\rho_t) \geq 0$ for all initial states $\rho$. Hence, the entropy may monotonically increase even for non-Markovian random unitary qubit dynamics.
Our discussion can be easily generalized for arbitrary $N>2$. We conjecture that  conditions (\ref{Ngamma}) are sufficient for the negative information flow. This point, however, deserves further investigation.

\vspace{.2cm}

\noindent This work was partially supported by the National Science Center project
DEC-2011/03/B/ST2/00136. We thank \'I\~nigo Luis Egusquiza for his remarks.



\begin{thebibliography}{1} \bibliographystyle{plain}

\bibitem{Breuer} H.-P. Breuer and F. Petruccione,
{\em The Theory of Open Quantum Systems} (Oxford Univ. Press,
Oxford, 2007).

\bibitem{GKS} V. Gorini, A. Kossakowski, and E. C. G. Sudarshan, J. Math. Phys.
{\bf 17}, 821 (1976).

\bibitem{Lindblad}  G. Lindblad, Comm. Math. Phys. {\bf 48}, 119
(1976).

\bibitem{Alicki} R. Alicki and K. Lendi, {\it Quantum Dynamical
Semigroups and Applications} (Springer, Berlin, 1987).



\bibitem{Strunz1} W. T. Strunz, L. Di\'osi, and N. Gisin, Phys. Rev. Lett. {\bf 82}, 1801
(1999);  Phys. Rev. A {\bf 58}, 1699 (1998).

\bibitem{nM} S. Daffer, K. W\'odkiewicz, J.D. Cresser, and J.K. McIver,
Phys. Rev. A {\bf 70}, 010304 (2004).

\bibitem{nM2} A. Shabani and D.A. Lidar, Phys. Rev. A {\bf 71}, 020101(R) (2005).

\bibitem{nM3} S. Maniscalco, Phys. Rev. A
{\bf 72}, 024103 (2005); {\em ibidem}. \textbf{75}, 062103 (2007); S. Maniscalco and F. Petruccione, Phys.
Rev. A {\bf 73}, 012111 (2006).

\bibitem{Wolf2} M.~M. Wolf and J.~I. Cirac, Comm. Math. Phys. \textbf{279}, 147 (2008); M. M. Wolf, J. Eisert, T. S. Cubitt and J. I. Cirac, Phys. Rev. Lett. \textbf{101}, 150402 (2008).

\bibitem{nM4}  H.-P. Breuer and B. Vacchini, Phys.
Rev. Lett. {\bf 101} (2008) 140402; Phys. Rev. E {\bf 79}, 041147
(2009).

\bibitem{PRL} D. Chru\'sci\'nski and A. Kossakowski, Phys. Rev. Lett. {\bf
104}, 070406 (2010).


\bibitem{Piilo}  J. Piilo, S. Maniscalco, K. H\"ark\"onen and K.-A. Suominen, Phys. Rev. Lett. {\bf 100}, 180402
(2008); Phys. Rev. A {\bf 79}, 062112 (2009).




\bibitem{versus1} P. Haikka, J. D. Cresser, and S. Maniscalco,  Phys. Rev. A {\bf 83}, 012112 (2011)

\bibitem{versus2} B. Vacchini, A. Smirne, E.-M. Laine, J. Piilo, and H.-P. Breuer,
New J. Phys. {\bf 13}, 093004 (2011).

\bibitem{versus3} D. Chru\'sci\'nski, Kossakowski and \'A. Rivas,  Phys. Rev. A {\bf 83}, 052128 (2011).
 

\bibitem{RHP} \'A. Rivas, S.F. Huelga, and M.B. Plenio, Phys. Rev. Lett. {\bf 105}, 050403
(2010).

\bibitem{BLP} H.-P. Breuer, E.-M. Laine, J. Piilo, Phys. Rev. Lett. {\bf 103},
210401 (2009).



\bibitem{inne1} X.-M. Lu, X. Wang, and C. P. Sun, Phys. Rev. A {\bf 82}, 042103
(2010).

\bibitem{inne2} A. K. Rajagopal, A. R. Usha Devi, and R. W. Rendell, Phys.
Rev. A {\bf 82}, 042107 (2010).

\bibitem{Apollaro} T. J. G. Apollaro, C. Di Franco, F. Plastina, and M.
Paternostro, Phys. Rev. A {\bf 83}, 032103 (2011).

\bibitem{Nature} B.-L. Liu, L. Li, Y.-F. Huang, C.-F. Li, G.-C. Guo, E.-M Laine, H.-P. Breuer and J. Piilo,  Nature Physics, {\bf 7}, 931 (2011).

\bibitem{singular} S. C. Hou, X. X. Yi, S. X. Yu, and C. H. Oh, Phys. Rev. A {\bf 83},
062115 (2011); Phys. Rev. A {\bf 86}, 012101 (2012).

\bibitem{inne4} S. Luo, S. Fu, and H. Song, Phys. Rev. A {\bf 86}, 044101 (2012).


\bibitem{Franco-PRA} R. Lo Franco, B. Bellomo, E. Andersson, and G. Compagno,  Phys. Rev. A {\bf 85}, 032318 (2012).


\bibitem{JPB} D. Chru\'sci\'nski and A. Kossakowski,  J. Phys. B: At. Mol. Opt. Phys. {\bf 45},  154002 (2012).


\bibitem{WIT} D. Chru\'sci\'nski and A, Kossakowski, {\em Witnessing non-Markovianity of quantum evolution}. arXiv:1210.8079.

\bibitem{HINDU} A. R. Usha Devi, A. K. Rajagopal, S. Shenoy, and R. W. Rendell, Journal of Quantum Information Science, {\bf 2}, 47 (2012).

\bibitem{Bassano} B. Vacchini, J. Phys. B: At. Mol. Opt. Phys. {\bf 45}, 154007 (2012).

\bibitem{Franco} R. Lo Franco, B. Bellomo, S. Maniscalco, and G. Compagno, Int. J.  Mod. Phys. B {\bf 27}, 1245053 (2013).

\end{thebibliography}
\end{document}